\documentclass[twoside]{article}
\usepackage{cite}
\usepackage{epsf}
\usepackage{equations}
\usepackage{epic}

\columnwidth\textwidth
\makeatletter
\@addtoreset{equation}{section}
\@addtoreset{footnote}{section}
\makeatother

\begin{document}

\thispagestyle{empty}
\parskip=12pt
\raggedbottom

\begin{flushright}
BUTP--96/12
\end{flushright}

\vspace*{2cm}

\begin{center}
{\LARGE New fixed point action for SU(3) lattice gauge theory}
\footnote{Work supported in part by Schweizerischer Nationalfonds}

\vspace*{1cm}

Marc Blatter and Ferenc Niedermayer\footnote{On leave from the 
Institute of Theoretical Physics, E\"otv\"os University, Budapest} \\
Institut f\"ur theoretische Physik \\
Universit\"at Bern \\
Sidlerstrasse 5, CH--3012 Bern, Switzerland

\vspace*{0.5cm}

{May 1996} 

\vspace*{3cm}

\nopagebreak[4]

\begin{abstract}
We present a new fixed point action for SU(3) lattice gauge theory,
which has --- compared to earlier published fixed point actions ---
shorter interaction range and smaller violations of rotational symmetry
in the static $q\bar{q}$-potential even at shortest distances.
\end{abstract}

\end{center}

\newpage
\setcounter{page}{1}

% ================================================================
\section{Introduction}
\label{se:intro}
% ================================================================
Lattice Monte Carlo simulations provide a powerful tool in studying
non-perturbative effects in QCD. Promising applications are, for
instance, hadron spectroscopy, high temperature thermodynamics or the
structure of the QCD vacuum.  However, using a space-time lattice as an
ultraviolet regulator introduces discretization errors.  A careful
analysis is required to extrapolate the Monte Carlo results to zero
lattice spacing recovering continuum physics. The simplest choice of the
lattice action corresponding to the given continuum theory produces, in
general, quite large lattice artifacts.   It is advisable to introduce
more complicated lattice actions which can significantly reduce such
discretization errors.

Wilson's renormalization group (RG) approach \cite{WILSON1,WILSON2}
offers a promising method for eliminating lattice artifacts in
asymptotically free field theories
\cite{HASENFRATZ1,HASENFRATZ2,DEGRAND3,DEGRAND1,DEGRAND2,DEGRAND4}.
Using actions on the renormalized trajectory (RT) of a given RG
transformation (RGT) one eliminates lattice artifacts in physical
quantities. These were called perfect actions \cite{HASENFRATZ1}.
The RT starts from a fixed point (FP) on the critical surface. 
The FP action is defined by retaining the functional form of 
the FP and using it also for finite couplings. 
In other words, the FP is a point in the space of
couplings (including the prefactor $\beta \propto 1/g^2$) while the FP
action is a straight line going in the direction of $g^2$).  The FP
action is a good approximation to the RT.  It has the important property
that it defines a perfect classical lattice theory: The solutions to the
lattice equations of motion have exactly the same value of the lattice
action as the continuum action on the corresponding continuum solutions.
For example, the FP as a classical action has scale invariant instanton
solutions when the corresponding continuum theory has.  Furthermore, it
can be argued that the RT coincides with the FP action even in 1-loop
perturbation theory --- the FP action is `1--loop perfect'
\cite{WILSON2,DEGRAND1,FARCHIONI}.  In the last years the FP action has
been constructed for several two-- and four dimensional asymptotically
free theories
\cite{HASENFRATZ1,HASENFRATZ2,DEGRAND3,DEGRAND1,DEGRAND2,DEGRAND4,FARCHIONI,WIESE,BIETENHOLZ,BLATTER,KUTI,BURKHALTER,KERRES,DEGRAND5}.

The type of blocking used in the RGT is to a large extent arbitrary.
But the properties of the FP, most notably the interaction range, are
very sensitive to the choice of the parameters in the blocking
transformation.  It has been demonstrated that one can choose the RGT in
such a way that the FP action is essentially concentrated on the unit
hyper-cube. This is important since one has to simplify the FP action in
order to use it in numerical simulations. In measuring physical
quantities, like masses, the string tension or critical temperature
approximate FP actions seems to give much reduced cut--off effects.  But
in expectation values of local operators --- like the static
$q\bar{q}$-potential --- new lattice artifacts appear. To reduce such
cut-off effects one has to use for a given FP action (or even for an
action on the RT) improved local operators.  In the RG approach the
origin of these artifacts is quite obvious.  Although the partition
function and all long range properties (masses, etc.) are unchanged by a
RG step, the fields on the coarse lattice represent some averages over
the original fine fields. If the blocking procedure violates rotational
invariance, correlation functions of the coarse fields will show a
violation of this symmetry as well. (As an analogy, in continuum
electrodynamics the potential between two square--shaped charge
distributions is not given by the $1/r$ potential alone, it has
contribution due to higher multipole moments as well.  Still the
physical content is described exactly, only the testing objects are
chosen inconveniently.) One way out, as mentioned above, is to construct
better fields (FP fields) --- another is to use a better, i.e. more
rotationally symmetric RGT to make the testing objects, the bare fields,
more spherical. The most important requirement remains, of course, the
short interaction range of the corresponding FP action. 

The present work has been motivated by the fact that the perturbative
potential of the {\it bare} Polyakov loops for the FP actions considered
in ref.~\cite{DEGRAND1,DEGRAND2} has shown a considerable violation of
rotational symmetry at a distance of 1--2 lattice units. Our aim is to
extend the class of blocking transformations to contain `more
rotationally invariant' ones.  By optimizing the free parameters of the
transformation for a short interaction range, we obtain a FP action
which has, as expected, also smaller violation of rotational symmetry in
the correlators of bare Polyakov loops.  Apart from the complications
connected with the more general form of the RGT this work repeats the
steps described in detail in ref.~\cite{DEGRAND1,DEGRAND2} and we shall
try to avoid unnecessary repetitions.  

While a parametrization of the FP action should represent the true FP
action sufficiently well, it should also be simple enough to be used in
numerical simulations.  Parametrization errors are one source of
`imperfectness' in the present approach. Another is that the FP action
is only an approximation --- although a very good one --- to the perfect
action, i.e.\ an action on the RT. The latter can be improved by trying
to follow the RT \cite{HASENFRATZ1,KUTI,GOTTLOB}, the former by
including more operators in the parametrization.  How far one should go
in this direction depends on many things --- how large are the remaining
lattice artifacts, what is the overhead for including further operators,
what is the gain in computer time by working on coarser lattices, etc. A
detailed study of the cut--off effects is needed to decide how a given
lattice action performs.  In this paper only the first few steps are
taken in this direction for the new RGT. 

The Wilson action in the SU(3) gauge theory gives huge cut-off effects
for thermodynamic quantities (e.g. about 50\% for $N_t = 4$,
$\beta\to\infty$), therefore it is especially important to use improved
actions for such studies \cite{BEINLICH,BOYD2}.  In a related paper
A.~Papa has studied lattice artifacts in thermodynamics with the new FP
action considered here \cite{PAPA}.

An alternative way of improving lattice actions is the Symanzik approach
\cite{SYMANZIK,LUESCHER1,WEISZ,LUESCHER2}.   Here one adds to the action
order by order in perturbation theory appropriate irrelevant operators
which cancel the leading cut--off effects. Although it is based on
perturbation theory, the idea can be extended beyond the perturbative
region as well \cite{JANSEN}.   The tadpole improvement also aims at
extending the Symanzik approach beyond the perturbative regime, using a
non--perturbative phenomenological determination of the coefficients of
the correction terms \cite{LEPAGE,ALFORD}.

We note here that the use of FP actions is especially promising in the
study of topological effects \cite{BLATTER,BURKHALTER,DEGRAND5,DELIA},
because in this approach there are no dangerous `dislocations' and an
improved definition of the topological charge can be given.

The paper is organized as follows:  First we review the RGT and its FP
with special consideration to the new block transformation.  We study
the quadratic expansion and determine the optimal block transformation
parameters. This is followed by the construction of three different loop
parametrizations of the FP action.  In section \ref{se:sqq} we study
cut-off effects in the quadratic approximation of the static
$q\bar{q}$-potential comparing the Wilson action with the FP actions
derived earlier and the new FP action. To set the physical scale we
measure the critical couplings for various lattice sizes.   This is done
in section \ref{sse:betac}.  We then study the performance of the new FP
action by evaluating the static $q\bar{q}$-potential using Polyakov loop
correlators.  Finally, in the last section we present some conclusions.

% ================================================================
\section{The FP action}
\label{se:fpact}
% ================================================================
With emphasis on the new block transformation we briefly review the RGT
and its FP presented in ref.~\cite{DEGRAND1}.  We evaluate the FP action
in a quadratic approximation and fix the free parameters in the block
transformation.  Finally, we construct simple parametrizations of the FP
action, one of which will be used in MC simulations.

We consider a SU(N) pure gauge theory\footnote{Although the numerical
studies have been performed for SU(3), the equations are mostly given
for general N.} in four dimensional Euclidean space defined on a
periodic lattice.  The partition function is defined as follows
\begin{equation} \label{eq:parti}
Z = \int \! DU \, \mbox{e}^{-\beta S(U)},
\end{equation}
where $DU$ is the invariant group measure and $\beta S(U)$ is some
lattice regularization of the continuum action.  Starting from  such a
lattice action we perform a real space RGT by defining a new action by
\begin{equation} \label{eq:rgact}
\mbox{e}^{-\beta' S'(V)} = \int DU 
\exp \,\left[ -\beta \left\{\,  S(U)+T(U,V) \, \right\} \right].
\end{equation}
Here $T(U,V)$ is the blocking kernel of the transformation and is
defined as
\begin{equation} \label{eq:truv}
T(U,V)= \sum_{n_B,\mu} \left( 
{\cal N}_{\mu}(n_B) - {\kappa \over N} \;
\mbox{Re Tr} \, \left[\, V_{\mu}(n_B) Q^{\dagger}_{\mu}(n_B) \,\right] 
 \right),
\end{equation}
where $Q_\mu(n_B)$ is a $N\times N$ matrix which represents some mean
of products of link variables $U_\mu(n)$ connecting the sites
$2n_B$ and $2n_B+2\hat{\mu}$ on the fine lattice.  The term ${\cal
N}_\mu(n_B)$ is a normalization which ensures the invariance of the
partition function.  The parameter $\kappa$ is a positive constant.
It will be fixed later to optimize the quadratic lattice action.  

The block transformation is defined by a prescription for constructing
the averaging matrices $Q_\mu(n_B)$. Two such block transformations were
defined in ref.~\cite{DEGRAND1}.  Here we introduce one, which we call
type~III blocking.  Instead of using just simple staples we also build
`diagonal staples' by going first in the planar or spatial diagonal
directions orthogonal to $\hat{\mu}$, then after a step in direction
$\hat{\mu}$ returning along the corresponding diagonal.  To be specific,
first we create the matrices $W^{(m)}(n,n')$ connecting the sites $n$
and $n'$, where $n'$ is a site with coordinates $n_{\mu}' - n_{\mu} =0$,
$|n_{\nu}' - n_{\nu}| \le 1$ (for any $\nu$): 
\begin{subequations}
\begin{eqnarray} \label{eq:type3ww} 
& & W^{(0)}(n,n) = 1 \, , 
                                           \label{eq:type3wa}    \\
& & W^{(1)}(n,n+\hat{\nu})            = U_{\nu}(n) \, , 
                                           \label{eq:type3wb}    \\
& & W^{(2)}(n,n+\hat{\nu}+\hat{\rho}) = \frac{1}{2}
\left( U_{\nu}(n)U_{\rho}(n+\hat{\nu}) + 
       U_{\rho}(n)U_{\nu}(n+\hat{\rho}) \right)\, ,
                                           \label{eq:type3wc}    \\
& & W^{(3)}(n,n+\hat{\nu}+\hat{\rho}+\hat{\lambda}) = 
\frac{1}{6} \left(
U_{\nu}(n)U_{\rho}(n+\hat{\nu}) U_{\lambda}(n+\hat{\nu}+\hat{\rho})
+ \mbox{permutations} \right) \, . 
                                           \label{eq:type3wd}
\end{eqnarray}
\end{subequations} 
Here $\nu$, $\rho$ and $\lambda$ go over all (positive and negative)
directions different from $\mu$ and from each other.
(Of course, $U_{-\nu}(n)=U_{\nu}^{\dagger}(n-\hat{\nu})$.
Values of $W^{(m)}(n,n')$ not indicated are taken to be zero.)
$W^{(2)}(n,n')$ represents the `planar diagonal link', 
$W^{(3)}(n,n')$ the spatial one. 
In eqs.~(\ref{eq:type3wc}), (\ref{eq:type3wd})
the sum is taken over all shortest paths leading to the 
endpoint $n'$ of the corresponding diagonal.
The fuzzy link operator is constructed then by a modified Swendsen 
smearing \cite{SWENDSEN}:
\begin{equation} \label{eq:type3w} 
\mathsf{W}_\mu(n) = \sum_{m=0}^{3} \sum_{n'} c_{m} W^{(m)}(n,n')
U_{\mu}(n') W^{(m)}(n'+\hat{\mu},n+\hat{\mu}).
\end{equation} 
The coefficients $c_m$ are free parameters subject to the constraint: 
\begin{equation} \label{eq:sumc}
c_0 + 6 \, c_1 + 12 \, c_2 + 8 \, c_3 = 1.
\end{equation}
Finally, the matrix $Q_\mu(n_B)$ is the product of two fuzzy link 
operators connecting the points $2n_B$ and $2n_B+2\hat{\mu}$ on the 
fine lattice:
\begin{equation} \label{eq:type3q}
Q_\mu(n_B) = \mathsf{W}_\mu(2n_B) \mathsf{W}_\mu(2n_B + \hat{\mu}).
\end{equation} 
The condition eq.~(\ref{eq:sumc}) ensures that for a trivial field 
configuration $Q_\mu(n_B)$ is equal to the unit matrix.

In the limit $\beta \rightarrow \infty$ eq.~(\ref{eq:rgact}) 
reduces to a saddle point problem.  At the FP this is an implicit
equation for the FP action $S^{FP}$:
\begin{equation} \label{eq:fpeq}
S^{FP}(V) = \min_{U} \left\{ S^{FP}(U) +  
 \sum_{n_B,\, \mu} \left( {\cal N}^\infty_\mu(n_B)
 - \frac{\kappa}{N} \; \mbox{Re Tr}\, \left[ V_\mu(n_B) 
 Q^\dagger_\mu(n_B) \right] \, \right) \; \right\}.
\end{equation}
The normalization constant at $\beta=\infty$ becomes
\begin{equation}
{\cal N}^\infty_\mu(n_B) = 
   \max_{ W \in \mbox{\scriptsize SU(N)} }
\left\{ \frac{\kappa}{N} \;
\mbox{Re Tr} \, [\, W Q^\dagger_\mu(n_B) \,] \right\}.
\end{equation}

As was already shown in ref.~\cite{HASENFRATZ1,DEGRAND1} the FP action 
has scale invariant instanton solutions.  This is a consequence of the 
FP equation (\ref{eq:fpeq}) alone and applies also for type~III 
blocking.  We will not proceed along these lines however, but will turn 
to an expansion in the vector potentials.

% ================================================================
\subsection{The quadratic approximation}
\label{sse:qaude}
% ================================================================
Again this section is very similar to the one presented in 
ref.~\cite{DEGRAND1}.  The only difference to type~I and type~II 
transformations is the actual transformation tensor involved.

The FP equation~(\ref{eq:fpeq}) is valid for arbitrary configurations
$\{V\}$.  For smooth configurations it can be expanded in powers of the 
vector potentials $B_\mu(n_B)$ and $A_\mu(n)$:
\begin{equation}
V_\mu(n_B) = e^{iB_\mu(n_B)}, \qquad U_\mu(n) = e^{iA_\mu(n)}.
\end{equation}

The most general form of the quadratic action can be written in Fourier 
space as
\begin{equation} \label{eq:quada}
 2N \, S(U) = \frac{1}{V} \sum_{k}
\tilde{\rho}_{\mu \nu}(k) \; \mbox{Tr}
\left[ \tilde{A}_{\mu}(-k)\tilde{A}_{\nu}(k) \right]
+{\rm O}\left( \tilde{A}^3\right),
\end{equation}
where $V$ is the volume of the fine lattice and
$\tilde{\rho}_{\mu \nu}(k)$ 
are the quadratic coefficients to be determined\footnote{In the 
following we suppress the tilde for Fourier transformed quantities.}.  
The transformation kernel becomes 
\begin{equation} \label{eq:quadt}
2N \, T(U,V) = \frac{\kappa}{V_B} \sum_{k_B}
      \mbox{Tr} \left[ \left( \Gamma_{\mu}(-k_B)-B_{\mu}(-k_B)\right)
                       \left( \Gamma_{\mu}(k_B)-B_{\mu}(k_B)\right)
		\right] 
 + {\rm O}\left( {\rm cubic} \right),
\end{equation}
where $V_B=V/16$ is the volume of the coarse lattice.
The matrices $\Gamma_{\mu}(k_B)$ represent the linear contributions 
to the averages $Q_\mu(n_B)$ and can be written as
\begin{equation} \label{eq:wmunu}
\Gamma_\mu(k_B) = \frac{1}{16}\sum_{l = 0}^{1} 
\omega_{\mu\nu}(\frac{k_B}{2}+\pi l) A_\nu(\frac{k_B}{2}+\pi l),
\end{equation}
where $l = (l_0,l_1,l_2,l_3)$ is an integer vector and the summation goes 
over $l_\mu = 0,\, 1$.
The tensor $\omega_{\mu \nu}$ is fixed by the form of the blocking
kernel:
\begin{equation}
\omega_{\mu \nu}(k)=\left( 1+{\rm e}^{ik_{\mu}}\right)
\left[  c_0 \delta_{\mu \nu} +
      6 c_1 \tau^{(1)}_{\mu \nu}(k) + 12 c_2 \tau^{(2)}_{\mu \nu}(k) +
      8 c_3 \tau^{(3)}_{\mu \nu}(k) 
\right].
\end{equation}
Here the coefficients $c_m$ were introduced in eq.~(\ref{eq:type3w}). 
The quantities $\tau^{(m)}_{\mu\nu}(k)$ are the linear 
contributions to the fuzzy link operator; they can be written as
\begin{eqnarray}
\tau^{(1)}_{\mu \nu}(k) & = & \frac{1}{6}
\left[ \;
\widehat{k}_{\mu} \widehat{k}_{\nu}^* + 
\delta_{\mu \nu}
  (6-\xi) 
\; \right], \nonumber \\
% <<<<<<<<<<<<<<<<<<<<<<<<<<<<<<<<<<<<<<<<<<<<<<<<<<<<<<<<<<<<<<
\tau^{(2)}_{\mu \nu}(k) & = & \frac{1}{24} 
\left[ \;
\widehat{k}_{\mu} \widehat{k}_{\nu}^* 
  (8 - \xi + \widehat{k}_{\mu} \widehat{k}_{\mu}^*
           + \widehat{k}_{\nu} \widehat{k}_{\nu}^* ) + 
\delta_{\mu \nu} 
  (24 - 8\xi+ \xi^2 - \eta - 
   \xi\widehat{k}_{\mu}\widehat{k}_{\mu}^*)
   \; \right], \nonumber\\
% <<<<<<<<<<<<<<<<<<<<<<<<<<<<<<<<<<<<<<<<<<<<<<<<<<<<<<<<<<<<<<
\tau^{(3)}_{\mu \nu}(k) & = & \frac{1}{48} 
\Big[ \;
\widehat{k}_{\mu} \widehat{k}_{\nu}^* 
  (24 - 6 \xi + \xi^2 - \eta 
    + 6 \widehat{k}_{\mu} \widehat{k}_{\mu}^*
    - 2 \xi \widehat{k}_{\mu} \widehat{k}_{\mu}^*
    + 2 ( \widehat{k}_{\mu} \widehat{k}_{\mu}^* )^2
    + 6 \widehat{k}_{\nu} \widehat{k}_{\nu}^*  
 \nonumber\\ & & ~~~~~~~~~~~~~
    - 2 \xi \widehat{k}_{\nu} \widehat{k}_{\nu}^* +
    2 ( \widehat{k}_{\nu} \widehat{k}_{\nu}^* )^2
    + 2 \widehat{k}_{\mu} \widehat{k}_{\mu}^*
        \widehat{k}_{\nu} \widehat{k}_{\nu}^* )  
                                                  \label{eq:tau}   \\
 & & ~~~~
+ \delta_{\mu \nu} 
  ( 48 -24\xi + 6 \xi^2 -\xi^3 
   -6\xi \widehat{k}_{\mu} \widehat{k}_{\mu}^*
   +2\xi^2\widehat{k}_{\mu} \widehat{k}_{\mu}^*
   -2\xi( \widehat{k}_{\mu} \widehat{k}_{\mu}^* )^2 
 \nonumber\\ & & ~~~~~~~~~~~~~~
  + 3\xi\eta - 2 \epsilon - 6 \eta - 
  2\eta \widehat{k}_{\mu} \widehat{k}_{\mu}^* )
\Big]. \nonumber
\end{eqnarray}
The lattice momentum $\widehat{k}_{\mu}$ is defined by
\begin{equation} 
\widehat{k}_{\mu} = \mbox{e}^{ik_\mu} - 1
\end{equation}
and we have introduced the following abbreviations
\begin{equation} \label{eq:abbr}
\xi \doteq |\hat{k}|^2 \doteq 
\sum_\mu \widehat{k}_{\mu} \widehat{k}_{\mu}^*,
\quad  
\eta  \doteq  \sum_\mu ( \widehat{k}_{\mu} \widehat{k}_{\mu}^* )^2, 
\quad 
\epsilon \doteq \sum_\mu ( \widehat{k}_{\mu} \widehat{k}_{\mu}^* )^3.
\nonumber
\end{equation}
As a consequence of gauge invariance $\tau^{(m)}_{\mu\nu}(k)$ and
$\omega_{\mu\nu}(k)$ satisfy the relations
\begin{equation} 
\tau^{(m)}_{\mu\nu}(k) \widehat{k}_{\nu} = \widehat{k}_{\mu} \,, \quad
\omega_{\mu\nu}(k) \widehat{k}_{\nu} = \widehat{2 k}_{\mu} \,.
\end{equation}

Combining the above expansions the FP equation has at the quadratic 
level the following form
\begin{eqnarray} \label{eq:quadrg}
\lefteqn{ 
{1 \over V_B} \sum_{k_B} \rho'_{\mu \nu}(k_B) \;
\mbox{Tr}\left[ B_{\mu}(-k_B)B_{\nu}(k_B) \right] = 
\min_{A}\Big\{  {1\over V} \sum_k \rho_{\mu \nu}(k) \; \mbox{Tr}
\left[ A_{\mu}(-k)A_{\nu}(k) \right]~~~~~ }  \\ 
& & ~~~~~~~~~~~~~~~~~ 
\mbox{\hspace*{0.7mm}} + {\kappa \over V_B} \sum_{k_B} \mbox{Tr}
\left[ \left( \Gamma_{\mu}(-k_B)-B_{\mu}(-k_B)\right)
\left( \Gamma_{\mu}(k_B)-B_{\mu}(k_B)\right) \right]
\Big\}.
\nonumber
\end{eqnarray}

To proceed further we have to introduce a temporary gauge fixing to be
able to invert the tensors $\rho_{\mu\nu}$ and $\rho'_{\mu\nu}$.
One obtains then from eq.~(\ref{eq:quadrg}) a recursion relation 
for the propagator 
\begin{equation}
D_{\mu\nu}(k) = \rho^{-1}_{\mu\nu}(k).
\end{equation}
A way to find the FP solution for the propagator is to start from the 
Wilson propagator with some gauge fixing parameter $\alpha$: 
\begin{equation} \label{eq:prop0}
D^{(0)}_{\mu\nu}(k) =
\frac{ \delta_{\mu\nu} }{|\hat{k}|^2} + \alpha 
\frac{\widehat{k}_{\mu} \widehat{k}_{\nu}^*}{|\hat{k}|^4},
\end{equation}
and iterate the propagator to the FP. (At the end we have to switch off
the gauge fixing in $\rho_{\mu\nu}$, which means taking the limit
$\alpha \rightarrow \infty$.)   Using gauge relations for 
$\omega_{\mu\nu}(k)$ one can show that starting from the standard 
propagator, after an arbitrary number of iterations, the propagator 
$D_{\mu\nu}(k)$ assumes the form:  
\begin{equation} 
D_{\mu\nu}(k) =
G_{\mu\nu}(k) + \alpha f(k) \widehat{k}_{\mu} \widehat{k}_{\nu}^*,
\end{equation} 
where $G_{\mu\nu}(k)$ and $f(k)$ are independent of $\alpha$.  
Under the RGT they are iterated as follows:
\begin{equation}
G'_{\mu\nu}(k_B) = 
 \frac{1}{16} \sum_{l = 0}^1 
     \left[ \omega(\frac{k_B}{2} + \pi l) G(\frac{k_B}{2} + \pi l)
       \omega^\dagger(\frac{k_B}{2} + \pi l)
     \right]_{\mu\nu} +
    \frac{1}{\kappa} \delta_{\mu\nu},
\end{equation}
and
\begin{equation}
f'(k_B) = \frac{1}{16} \sum_{l = 0}^1  f(\frac{k_B}{2} + \pi l).
\end{equation}
For finite $\alpha$ the action density becomes
\begin{equation}
\rho_{\mu\nu}(k) = G^{-1}_{\mu\nu}(k) - \alpha f(k)
\frac{G^{-1}_{\mu\rho}(k)\widehat{k}_{\rho} \cdot
\widehat{k}_{\sigma}^*G^{-1}_{\sigma\nu}(k)}
{ 1 + \alpha f(k) \,
u(k)},
\end{equation}
where we have introduced
\begin{equation} \label{eq:uofk}
u(k) = \widehat{k}_{\sigma}^*G^{-1}_{\sigma\rho}(k)\widehat{k}_{\rho}.
\end{equation} 
In the limit $\alpha \rightarrow \infty$ one obtains
\begin{equation}
\rho_{\mu\nu}(k) = G^{-1}_{\mu\nu}(k) -
\frac{G^{-1}_{\mu\rho}(k)\widehat{k}_{\rho} \cdot
\widehat{k}_{\sigma}^*G^{-1}_{\sigma\nu}(k)}
{ u(k) }.
\end{equation}

The type~III RGT has four real parameters.  These are $c_1$, $c_2$, 
$c_3$ and  $\kappa$ in eqs.~(\ref{eq:type3w}) and (\ref{eq:truv}).
We tune them for a short ranged quadratic FP action.
Since we optimize in a larger set of blocking transformations 
(note that at the quadratic level the blocking transformation 
includes those considered in type~I blocking),  the final 
action is expected to have a shorter interaction range. 
We found the following parameters to be optimal:
\begin{equation}
c_1 = 0.07, \quad c_2 = 0.016, \quad c_3 = 0.008, \quad \kappa = 8.8.
\end{equation}
The iteration procedure converges quite rapidly with a next-to-leading
eigenvalue of 0.25.
While for type~I and type~II blocking the largest  couplings 
of $\rho^{FP}_{\mu\nu}(r)$ decrease as $\exp(-2.7r)$, 
rsp.\ $\exp(-3.1r)$, for type~III blocking it is $\exp(-3.4r)$.  
Some of the values of $\rho^{FP}_{\mu\nu}(r)$ have been
listed in table \ref{tab:rho}.  

\begin{table}
\begin{center}
\begin{tabular}{r r|r r}      
%
%  based on:
%  akappa=    8.800  c=   0.070000  0.017000  0.006000 alpha=       0.0  
% 
\hline
  $r$~~~~     &    $\rho_{00}(r)$~  & $r$~~~~ &  $\rho_{10}(r)$~ \\   
\hline
0  0  0  1  & -0.47938   &     0  0  0  0  & -0.72414 \\
0  0  1  1  & -0.10005   &     0  0  0  1  & -0.05969 \\
0  1  1  1  & -0.03110   &     0  0  1  1  & -0.00991 \\
1  0  0  1  & ~0.00588   &     2  1  0  0  & -0.00222 \\
1  0  1  1  & -0.00457   &     0  1  0  1  & -0.00204 \\
1  1  1  1  & -0.00316   &     2  1  0  1  & -0.00081 \\
1  0  0  2  & ~0.00208   &     0  1  1  1  & -0.00059 \\
1  0  0  0  & ~0.00204   &     2  1  1  1  & -0.00043 \\
2  0  0  0  & ~0.00138   &     0  2  0  0  & -0.00023 \\
0  0  0  2  & ~0.00132   &     0  0  0  2  & ~0.00022 \\
0  0  1  2  & -0.00084   &     0  0  1  2  & -0.00011 \\
\hline
\end{tabular}
\end{center}
\caption{Some of the elements of $\rho_{00}(r)$ and $\rho_{10}(r)$ in
configuration space for the RGT of type~III.  
The values are obtained after $5$ RG steps on a $16^4$ lattice.}
\label{tab:rho}
\end{table}

Solving eq.~(\ref{eq:quadrg}) we obtain a linear relation between the 
field $B_\mu$ on the coarse lattice and the minimizing field $A_\mu$
on the fine lattice:
\begin{equation} \label{eq:zdef}
A_{\mu}(k) = Z_{\mu\nu}(k) B_{\nu}(2k),
\end{equation}
where
\begin{equation}
Z_{\mu\nu}(k) = \left[ D(k) \omega^\dagger(k) D'^{-1}(2k) \right]_{\mu\nu}.
\end{equation}
The connecting tensor $Z_{\mu\nu}$ enters, in particular, 
in the construction of FP operators \cite{DEGRAND1}.
For large $\alpha$ it is given by:
\begin{eqnarray}
Z_{\mu\nu}(k) & =  & \left[ G(k) \omega^{\dagger}(k)
G'^{-1}(2k) \right]_{\mu\nu} + \frac{1}{u'(2k)} \widehat{k}_{\mu}\cdot
\widehat{2k}_{\sigma}^* \left[ G'^{-1}(2k)\right]_{\sigma\nu} \\ & &
- \frac{1}{u'(2k)} 
 \left[ G(k) \omega^{\dagger}(k) G'^{-1}(2k) \right]_{\mu\sigma}
\widehat{2k}_{\sigma} \cdot \widehat{2k}_{\rho}^* \left[ G'^{-1}(2k)
\right]_{\rho\nu} + \mbox{O}\left(\frac{1}{\alpha}\right). 
\nonumber
\end{eqnarray} 
where $u'(k)$ is defined in eq.~(\ref{eq:uofk}) with $G$ replaced by
$G'$.  The gauge relation for $Z_{\mu\nu}$ reads
\begin{equation}
Z_{\mu\nu}(k) \widehat{2k}_{\nu} = \widehat{k}_{\mu}.
\end{equation}
Some of the $Z_{\mu\nu}(r)$ values are listed in table~\ref{tab:Z}.  

\begin{table}
\begin{center}
\begin{tabular}{r r|r r}
\hline
 $r$~~~~ & $Z_{00}(r)~$ &$r$~~~~& $Z_{10}(r)$~  \\  
\hline
  0  0  0  1  &  0.12507   &   0  1  0  0  &  -0.03993 \\
  0  0  1  1  &  0.06061   &  -1  1  0  0  &  -0.01336 \\
 -1  0  0  0  &  0.05378   &  -1  0  0  0  &  -0.01237 \\
 -1  0  0  1  &  0.04776   &   0  1  1  0  &  -0.01164 \\
  0  1  1  1  &  0.03794   &   0  2  0  0  &  -0.01053 \\
 -1  0  1  1  &  0.02692   &   0  2  0  1  &  -0.00485 \\
 -2  0  0  0  &  0.02274   &  -1  1  0  1  &  -0.00484 \\
 -1  1  1  1  &  0.01785   &   0  1  0  2  &  -0.00451 \\
  0  0  0  2  &  0.01165   &   0  1  1  1  &  -0.00416 \\
 -2  0  0  1  &  0.01058   &  -1  2  0  0  &  -0.00409 \\
 -1  0  0  2  & -0.00963   &   0  0  0  1  &  -0.00406 \\
\hline
\end{tabular}
\end{center}
\caption{Some of the elements of the tensor $Z_{\mu \nu}(r)$ in
configuration space. The lattice size is $16^4$ and 5 RG steps were
taken.}
\label{tab:Z}
\end{table}

% ================================================================
\subsection{Beyond the quadratic approximation}
\label{sse:loop}
% ================================================================
For the use in numerical simulations we need a parametrization of the FP
action $S^{FP}$ in terms of gauge invariant products of link
variables $U_\mu(n)$. For type~I and type~II RGT this has been studied
in detail in ref.~\cite{DEGRAND2}.  Here we shall simply outline the
procedure and present the results.

We parametrize the FP action by powers of traces of loops:
\begin{equation} \label{eq:parl}
S(U) = \frac{1}{N} \sum_{{\cal C}} \, \sum_{m} 
% c_m({\cal C}) \left( N - \mbox{Re Tr}\,( U_{{\cal C}}) \right)^m,
 c_m({\cal C})\; \left( \mbox{Re Tr}\,
 [ 1 - U_{{\cal C}} ]\right)^m,
\end{equation}
where $U_{{\cal C}}$ denotes the product of link variables $U_\mu(n)$
along the closed path ${\cal C}$. 
For a set of quadratically independent loops the coefficients 
$c_1({\cal C})$ can be obtained from the quadratic approximation 
$\rho_{\mu\nu}(r)$. We apply a numerical procedure to determine 
those coefficients with $m\ge2$. We generated about $500$ 
configurations $V$ using the Wilson action
with $\beta$ ranging from $5.1$ to $50.0$ and determined the
corresponding fine configurations $U(V)$ by numerical minimization. The
procedure is simplified by the observation that the typical value of the
action density on the minimizing configuration $U(V)$ in
eq.~(\ref{eq:fpeq}) is by a factor 30--40 smaller than the action
density on the coarse configuration $V$. This allows us two make a two
step iteration: First by using a good quadratic approximation on the
fine lattice one determines a parametrization of the l.h.s.\ which
includes higher powers and describes the `measured' values of the action
well (i.e.\ those obtained by the minimization procedure). In the next
step this precise intermediate parametrization is used on the
r.h.s.\ of eq.~(\ref{eq:fpeq}) to determine the value $S^{FP}(V)$ for
any configuration $V$, including the very rough ones.  The final step is
to represent the set $\{V^{(i)},\;S^{FP}(V^{(i)})\}$ by some form
suitable for numerical simulations.  Considering a larger number of
loops in eq.~(\ref{eq:parl}) one gets a better parametrization, but the
computer time grows rapidly with the length of the loops, hence one has
to restrict the number of loops.   

Below we present three  alternative parametrizations
of the FP action of type~III blocking.  They are denoted as type IIIa,
IIIb, and IIIc.  As has been mentioned earlier
\cite{HASENFRATZ1,DEGRAND1} there are many parametrizations which
represent the given set of action values equally well.  The optimal
values are not unique --- the matrix in the corresponding $\chi^2$ fit
has many nearly zero eigenvalues.
Since the roughest configurations dominate the fit our results 
will represent the FP action on typical configurations generated 
with $\beta_{\mbox{\scriptsize Wilson}} \sim 5-6$. 

\begin{figure}[htbp]
\begin{minipage}[t]{44mm}
\begin{center}
\begin{picture}(50,50)(0,0)
\put(0,0){\circle*{4}}
\put(0,0){\line(1,0){40}}
\put(40,0){\circle*{4}}
\put(40,0){\line(0,1){40}}
\put(40,40){\circle*{4}}
\put(40,40){\line(-1,0){40}}
\put(0,40){\circle*{4}}
\put(0,40){\line(0,-1){40}}
\end{picture}\\*[6mm]

pl
\end{center}
\end{minipage}
\begin{minipage}[t]{44mm}
\centering
\begin{picture}(90,50)(0,0)
\put(0,0){\circle*{4}}
\put(0,0){\line(1,0){40}}
\put(40,0){\circle*{4}}
\put(40,0){\line(1,0){40}}
\put(80,0){\circle*{4}}
\put(80,0){\line(0,1){40}}
\put(80,40){\circle*{4}}
\put(80,40){\line(-1,0){40}}
\put(40,40){\circle*{4}}
\put(40,40){\line(-1,0){40}}
\put(0,40){\circle*{4}}
\put(0,40){\line(0,-1){40}}
\dashline[30]{3}(40,0)(40,40)
\end{picture}\\*[6mm]

rt
\end{minipage}
\begin{minipage}[t]{54mm}
\centering
\begin{picture}(84,76)(0,0)
\put(0,0){\circle*{4}}
\put(0,0){\line(1,0){40}}
\put(40,0){\circle*{4}}
\put(40,0){\line(3,2){26}}
\put(64,16){\circle*{4}}
\put(64,16){\line(0,1){40}}
\put(64,56){\circle*{4}}
\put(64,56){\line(-1,0){40}}
\put(24,56){\circle*{4}}
\put(24,56){\line(-3,-2){26}}
\put(0,40){\circle*{4}}
\put(0,40){\line(0,-1){40}}
% hidden lines 
\dashline[30]{3}(0,40)(40,40)
\dashline[30]{3}(40,40)(64,56)
\dashline[30]{3}(40,0)(40,40)
\end{picture}\\*[6mm]

pg
\end{minipage}
\caption{Three different loops considered in the parametrization of the
FP action. The plaquette operator (pl), the rectangle (rt)
and the three-dimensional parallelogram (pg).}
\label{fig:loop}
\end{figure}
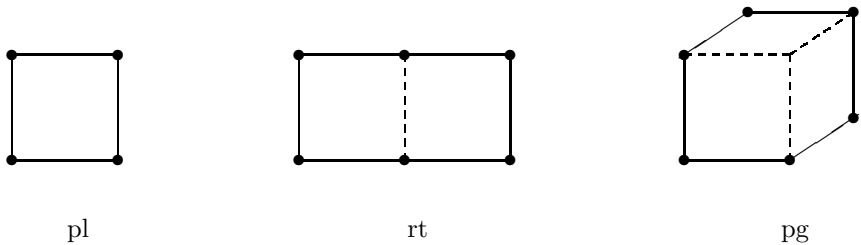

We shall consider the three loops shown in fig.~\ref{fig:loop}. 
In the parametrizations IIIa and IIIb we have chosen the same
set of operators as in ref.~\cite{DEGRAND2} for the type~I FP action,
the plaquette and the rectangle.
For type~IIIc action we have also included the rectangle.
The maximal power $m$ in eq.~(\ref{eq:parl}) for all loops was 4.
For the type~IIIb and type~IIIc parametrization we have used the 
freedom in the coefficients to demand that they satisfy the tree 
level Symanzik condition for spectral quantities \cite{WEISZ}:
\begin{equation} \label{eq:sym}
c_1(\mbox{pl}) + 20 c_1(\mbox{rt}) - 4 c_1(\mbox{pg}) = 0\, .  
\end{equation}
The results are listed in table \ref{tab:para}.
\begin{table*}[hbt]
% space before first and after last column: 1.5pc
% space between columns: 3.0pc (twice the above)
\setlength{\tabcolsep}{1.5pc}
\begin{tabular*}{\textwidth}{@{}l@{\extracolsep{\fill}}cllll}
% \begin{tabular*}{}{lcllll}
\hline
     &    operator & ~~~$c_1$  & ~~~$c_2$ & ~~~$c_3$  & ~~~$c_4$ \\  
\hline
IIIa & $\mbox{pl}$ & ~0.4822   & ~0.2288  & -0.1248   & ~0.02282 \\
     & $\mbox{pg}$ & ~0.06473  & -0.02245 & ~0.002954 & ~0.003471 \\
\hline
IIIb & $\mbox{pl}$ & ~0.3333   & ~0.4845  & -0.2730   & ~0.04858 \\
     & $\mbox{pg}$ & ~0.083333 & -0.01644 & -0.009525 & ~0.006324 \\
\hline
IIIc & $\mbox{pl}$ &  ~0.4792  & ~0.2226  & -0.1273   & ~0.02403 \\
     & $\mbox{rt}$ &  -0.0091  & -0.04471 & ~0.02563  & -0.003698 \\
     & $\mbox{pg}$ &  ~0.0742  & ~0.02047 & -0.02398  & ~0.007730 \\
\hline
\end{tabular*}
\caption{Couplings of the 8 parameter FP action of type~IIIa and
type~IIIb, and the 12 parameter FP action of type~IIIc.}
\label{tab:para}
\end{table*}
As stated above all the three fits represent the FP values of our set of
configurations equally well.  The deviations are similar to those shown
in ref.~\cite{DEGRAND2}.
We expect the three types of parametrizations IIIa, IIIb and IIIc
to give comparable results for simulations in the range 
$\beta_{\mbox{\scriptsize Wilson}}\sim5-6$.  
At small lattice spacings, however, when the linear approximation
dominates, the Symanzik condition imposed on type~IIIb and type~IIIc 
actions will eliminate the O($a^2$) effects, which makes them better
suited for perturbative calculations.
The couplings in table~\ref{tab:para} also satisfy
the normalization condition
\begin{equation} \label{eq:norm}
c_1(\mbox{pl}) + 8 c_1(\mbox{rt}) + 8 c_1(\mbox{pg}) = 1 \, .  
\end{equation}

Note that the whole quadratic part in the full FP action
satisfies exactly the Symanzik condition.  However,
in a parametrization using a restricted set of loops this condition may
not be satisfied any more --- unless it is specially
requested  as an auxiliary condition.  The quadratic part (the
coefficients $c_1({\cal C})$) used in the parametrization IIIc are
obtained from the best fit to the quadratic coefficients
$\rho_{\mu\nu}(r)$ --- with those operators under the constraint
(\ref{eq:sym}).  All the numerical simulations have been performed using
the parametrization IIIa.  The other two forms are given only to show
the ambiguity of the truncation step and to provide a parametrization
which maintains the Symanzik improved nature of the full FP action.

% ================================================================
\section{The static $q\bar{q}$-potential}
\label{se:sqq}  
% ================================================================
A good check for discretization errors is offered by the static
$q\bar{q}$-potential.  Below the critical temperature $T_c$ the
potential is expected to rise linearly for larger distances, with the
string tension as its slope.  For smaller distances the potential
should exhibit a Coulomb potential.  In lowest order
perturbation theory it is proportional to the time average of the zero
component of the free field propagator.  In the following two sections
we study the perturbative potential to lowest order and the full
potential using Polyakov loop operators. 

% =====================================================================
\subsection{The tree level potential}
\label{sse:qbq}
\begin{figure}[htb]
\begin{center}
%\vskip 10mm
\leavevmode
\epsfxsize=130mm
\epsfbox{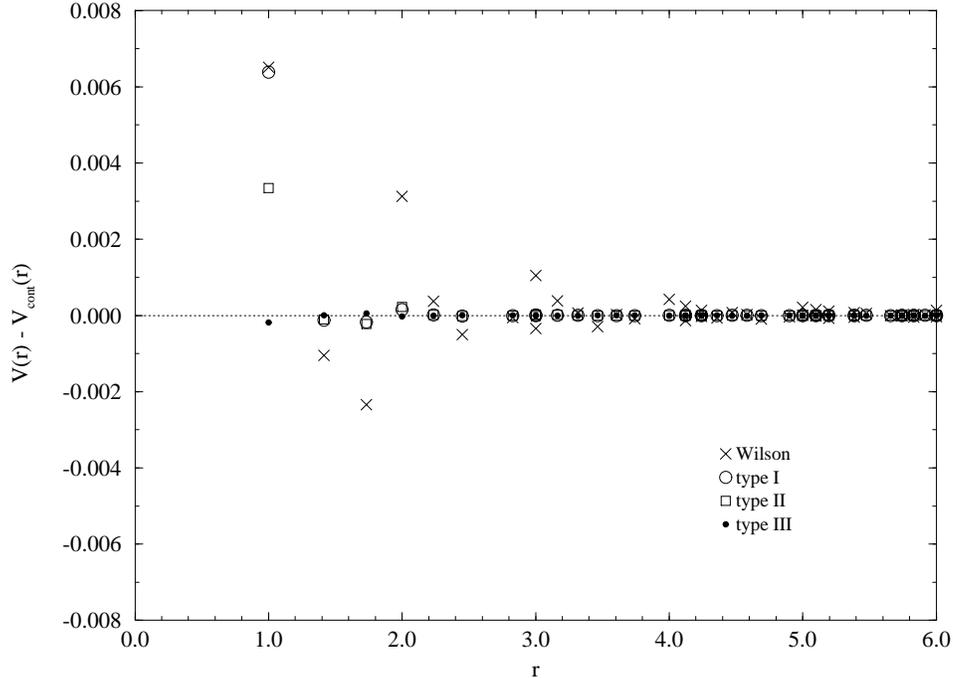}
%\vskip 10mm
\end{center}
\caption{ A comparison of the quadratic potential for different
blocking prescriptions in a cubic box of size $16^3$.  A non-physical
constant was subtracted to match the curves.}
\label{fig:quadpot}
\end{figure}

% =====================================================================
At the quadratic level the static potential in a finite continuum
box\footnote{Note that the force at $|\vec{x}|/L \approx 1/3$ in a finite box 
differs  from the infinite volume result by as much as 25\% so it is 
essential to use the correct finite box expression for comparison 
to the lattice results.}
of size $L^3_s$ is proportional to the Coulomb potential:
\begin{equation}
\label{eq:cbox}
V_{\mbox{\scriptsize cont}}(\vec{x}) = \frac{1}{L^3_s}
 \sum_{\vec{k} \neq 0}  \mbox{e}^{i\vec{k}\vec{x}} \frac{1}{(\vec{k})^{2}}, 
 \quad \mbox{where} \quad k_i = \frac{2 \pi}{L_s} l_i. 
\end{equation}
On a lattice of size $N^3_s$ and an arbitrary lattice action we get 
\begin{equation}
V(\vec{r}) = \frac{1}{N_s^3}
 \sum_{\vec{k} \neq 0}  \mbox{e}^{i\vec{k}\vec{r}} 
 D_{00}(k_0=0,\vec{k}), 
\end{equation}
where $D_{\mu\nu}(k)$ is the propagator corresponding to the quadratic
part of the lattice action.  
On fig.~\ref{fig:quadpot} we plotted the differences between the
lattice potentials and the continuum potential for Wilson action
and type I, II, and III actions.
To see how large this deviations are one can compare the force
measured on the lattice to the continuum force. 
In table~\ref{tab:force} the relative differences
\begin{equation}
\label{eq:error}
R(\vec{r}_1,\vec{r}_2)= {
\left( V(\vec{r}_1)
        -V_{\mbox{\scriptsize cont}}(\vec{r}_1) \right)
- \left( V(\vec{r}_2)
        -V_{\mbox{\scriptsize cont}}(\vec{r}_2) \right) \over
         V_{\mbox{\scriptsize cont}}(\vec{r}_1)
        -V_{\mbox{\scriptsize cont}}(\vec{r}_2) }
\end{equation}
are listed for the few first distances on the lattice.
\begin{table*}[hbt]
\setlength{\tabcolsep}{1.5pc}
\begin{tabular*}{\textwidth}{cccccc}
\hline
$\vec{r}_1$ & $\vec{r}_2$ & Wilson & type~I & type~II & type~III \\
\hline
% (1,1,0)    &  (1,0,0)   & ~0.33  & ~0.28~ &  ~0.15  & -0.008   \\
% (1,1,1)    &  (1,1,0)   & ~0.13  & ~0.005 &  ~0.01  & -0.004   \\
% (2,0,0)    &  (1,1,1)   & -0.90  & -0.05~ &  -0.07  & -0.012   \\ 
  (0,1,1) & (0,0,1) & ~0.325~ &  ~0.2797 & ~0.1479 &  -0.0080  \\
  (1,1,1) & (0,1,1) & ~0.125~ &  ~0.0051 & ~0.0113 &  -0.0044  \\
  (0,0,2) & (1,1,1) & -0.895~ &  -0.0544 & -0.0725 &  ~0.0125  \\
  (0,1,2) & (0,0,2) & ~0.662~ &  ~0.0034 & ~0.0493 &  -0.0048  \\
  (1,1,2) & (0,1,2) & ~0.283~ &  ~0.0092 & ~0.0118 &  -0.0022  \\
  (0,2,2) & (1,1,2) & -0.103~ &  -0.0016 & -0.0022 &  ~0.0004  \\
  (1,2,2) & (0,2,2) & ~0.183~ &  ~0.0017 & ~0.0023 &  -0.0005  \\
\hline
\end{tabular*}
\caption{The relative error of the perturbative force on the lattice
(cf. eq.~({\protect\ref{eq:error}})) for various lattice actions.}
\label{tab:force}
\end{table*}
There are two types of
deviations from the continuum result for the FP action \cite{DEGRAND1}.
The first type is due to the deviation of the shape of the blocking 
from a spherical averaging.  
This part contains, for example, a term $\propto P_4(\cos\theta)/r^5$ 
(an octupole term) and can be corrected by using FP Polyakov
loops\footnote{We have checked this explicitly by constructing FP
Polyakov loops in the linear approximation.}. 
The second type is an exponentially falling correction 
$\propto\exp(-r/r_0)$ where $r_0$ is of the order of the interaction
range and is due to the effect of the quantum fluctuations on the 
fine lattice. As seen from fig.~\ref{fig:quadpot}, for type~I and 
type~II actions the discretization errors are much reduced, but 
at $r=1$ a substantial deviation from the continuum can be observed. 
On the other hand, for type~III action the perturbative potential 
lies for all distances very near to the continuum result.

% ================================================================
\subsection{The temperature scale}
\label{sse:betac} 
% ================================================================
In this section we determine the critical couplings
$\beta_{c}$ for the colour deconfinement phase transition on lattices 
with different temporal extensions $N_t$.
As in ref.~\cite{DEGRAND2} we use the definition of the Columbia group 
\cite{CHRIST1}:  For a given lattice 
size $N^3_s\times N_t$ we perform MC simulations for various 
couplings $\beta$ around the expected transition region measuring the
Polyakov loop averaged over the lattice:
\begin{equation}
P = \frac{1}{N^3_s} \sum_{\vec{n}} \,
\mbox{Tr} \, \prod_{n_0=0}^{N_t-1} U_0(n_0,\vec{n})
\end{equation}
We evaluate the angle $\theta$ from 
$\tan(\theta)=\mbox{Im}(P)/\mbox{Re}(P)$ and measure the fraction 
of the time $f_{20}$ that $\theta$ lies in the range of 
$\pm 20^{\circ}$ of the directions of the Z(3) roots.  
Finally, the phase transition is defined as the coupling $\beta$ 
where the simulations yield for the deconfinement fraction
\begin{equation}
f_d = \frac{3}{2} f_{20} - \frac{1}{2}
\end{equation}
the value $0.5$.  

\begin{figure}[htb]
\begin{center}
\leavevmode
\epsfxsize=130mm
\epsfbox{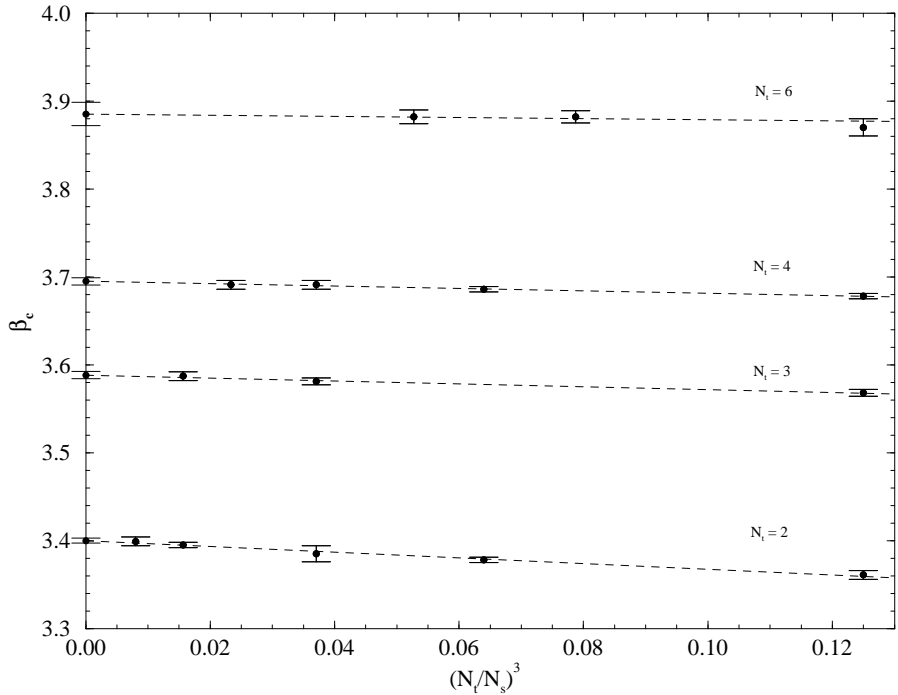}
\end{center}
\caption{Finite size scaling for the critical couplings of the
deconfinement phase transition using the parametrization IIIa.  
The $\beta=\infty$ extrapolations are included.}
\label{fig:fscaling}
\end{figure}

We measured the critical couplings for lattices of temporal extensions 
$N_t = 2$, $3$, $4$ and $6$ with different spatial sizes $N_s$. 
The results are listed in table \ref{tab:betacrit}.  The deconfinement
fraction has been determined at various couplings in the vicinity of the
phase transition.  At each coupling we have performed from $2000$ to
$30000$ sweeps.  The critical coupling $\beta_c$ was obtained by a
linear fit of the fractions bracketing $f_d(\beta_c)=0.5$.  Its error
was obtained by a jack-knife analysis.
\begin{table*}[hbt]
% space before first and after last column: 1.5pc
% space between columns: 3.0pc (twice the above)
\setlength{\tabcolsep}{1.5pc}
\begin{tabular*}{\textwidth}{@{}c@{\extracolsep{\fill}}cccc}
\hline
volume   & $N_t=2$     & $N_t=3$    & $N_t=4$ & $N_t=6$ \\
\hline
$4^3$    & 3.361(5)    &            &            &    \\
$5^3$    & 3.378(3)    &            &            &    \\
$6^3$    & 3.385(9)    & 3.568(4)   &            &    \\
$8^3$    & 3.395(3)    &            &  3.678(3)  &    \\
$9^3$    & 3.399(5)    & 3.581(4)   &            &    \\
$10^3$   &             &            & 3.686(3)   & 3.91(4)~~ \\
$12^3$   &             & 3.587(5)   & 3.691(5)   & 3.87(1)~~ \\
$14^3$   &             &            & 3.691(5)   & 3.882(7)~ \\
$16^3$   &             &            &            & 3.882(8)~ \\
$\infty$ &  3.400(3)   & 3.588(4)   & 3.695(4)   & 3.886(13) \\
\hline
\end{tabular*}
\caption{Critical couplings at finite volume and extrapolated to
infinite volume for the FP action with parameters IIIa in 
Table~{\protect\ref{tab:para}}.}
\label{tab:betacrit}
\end{table*}
The critical couplings in the thermodynamic limit ($N_s \rightarrow 
\infty$) were obtained by extrapolating the finite couplings using 
the finite size law \cite{CHRIST1,BOYD2}:
\begin{equation}
\beta_c(N_t,N_s)=\beta_c(N_t,\infty) - c(N_t) \left( 
\frac{N_t}{N_s} \right)^{3}.
\end{equation}
The finite size scaling behaviour is plotted in fig.\ \ref{fig:fscaling}.

% =========================================================================
\subsection{The full potential}
\label{sse:fpot}
\begin{figure}[htbp]
% \begin{center}
\noindent
\begin{minipage}[t]{75mm}
\centering
\epsfxsize=75mm
\epsfbox{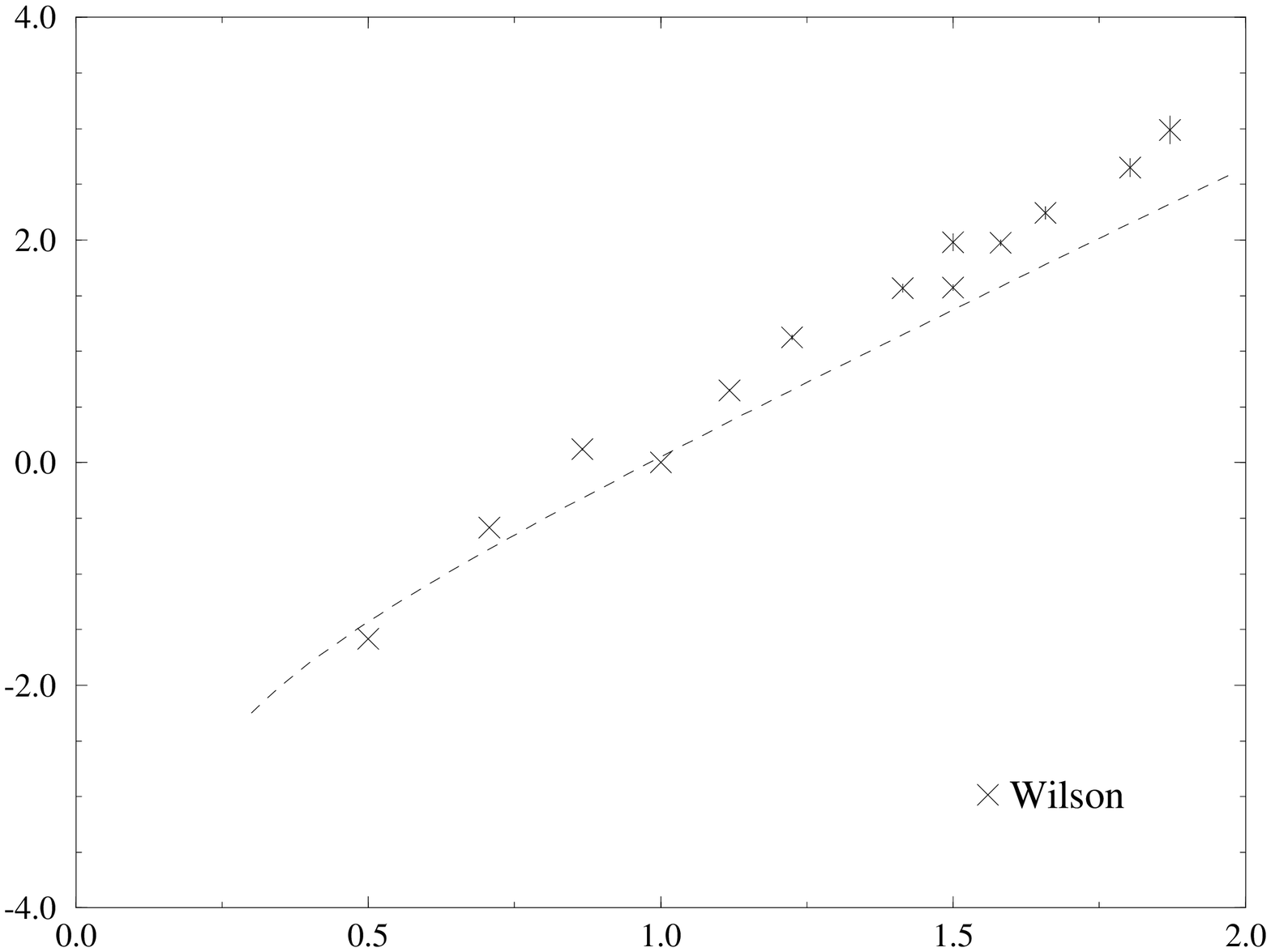}
\end{minipage}
\hfill
\begin{minipage}[t]{75mm}
\centering
\epsfxsize=75mm
\epsfbox{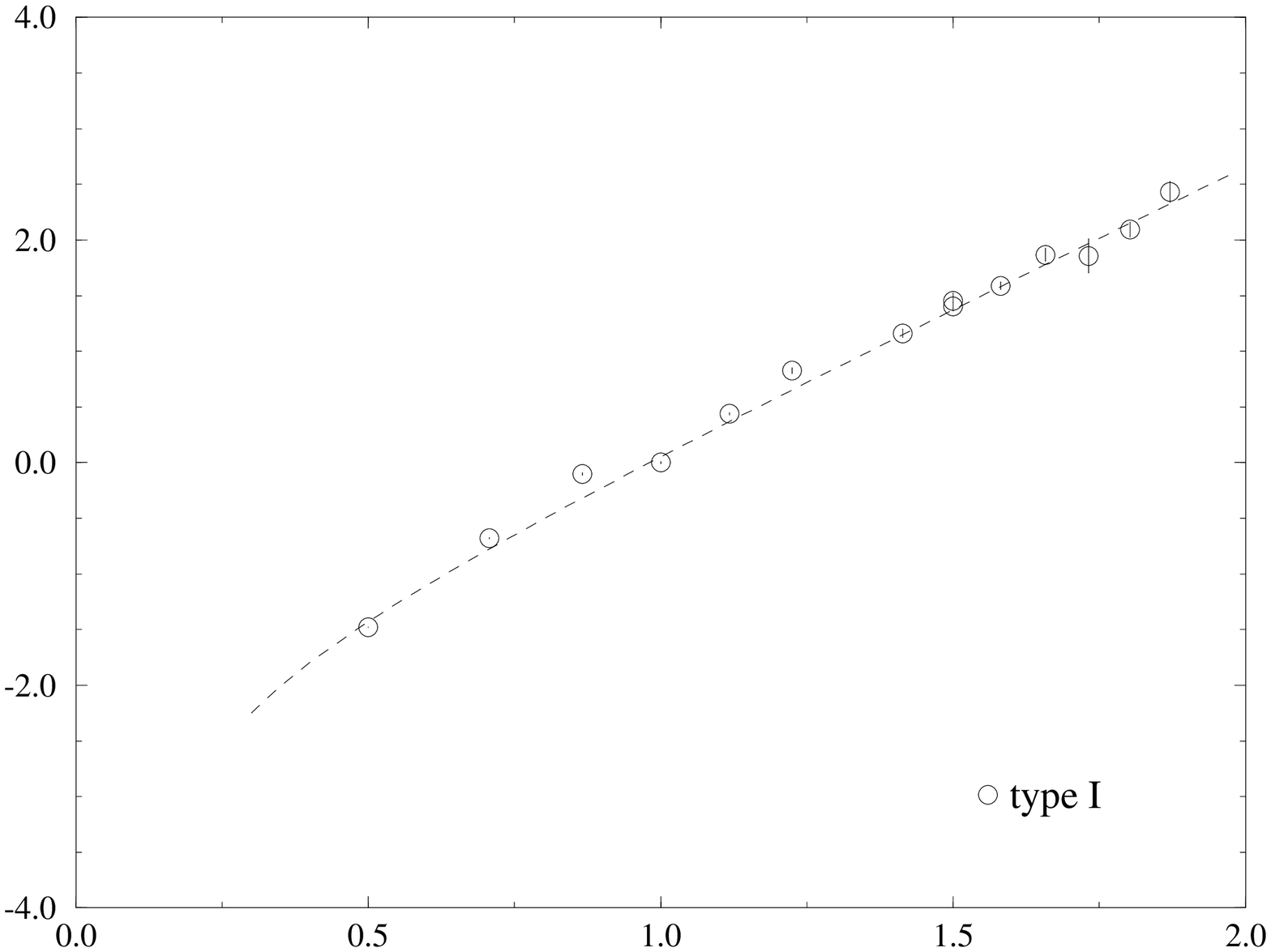}
\end{minipage} 
\begin{minipage}[t]{75mm}
\centering
\epsfxsize=75mm
\epsfbox{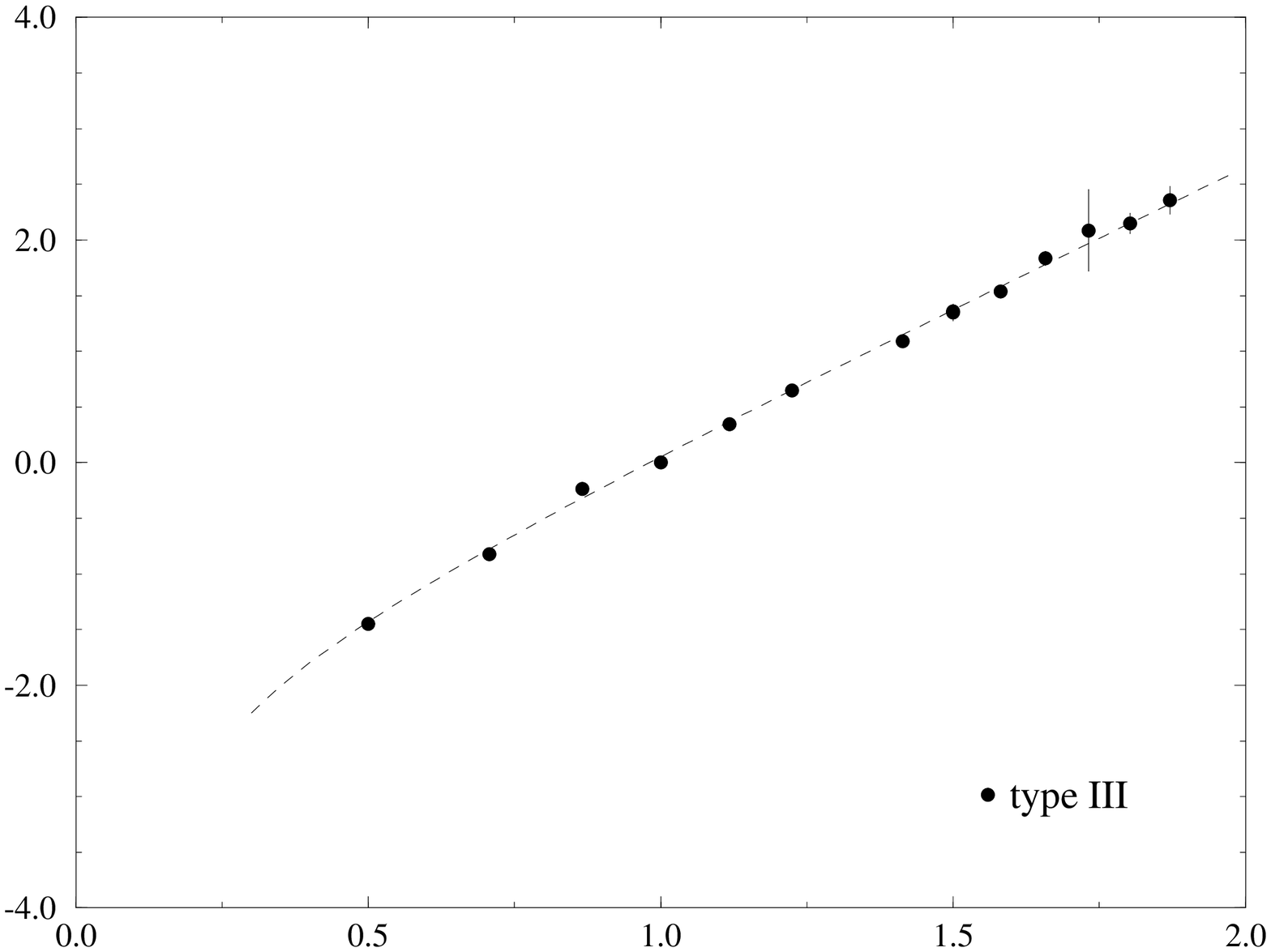}
\end{minipage}
\hfill
\begin{minipage}[t]{75mm}
\centering
\epsfxsize=75mm
\epsfbox{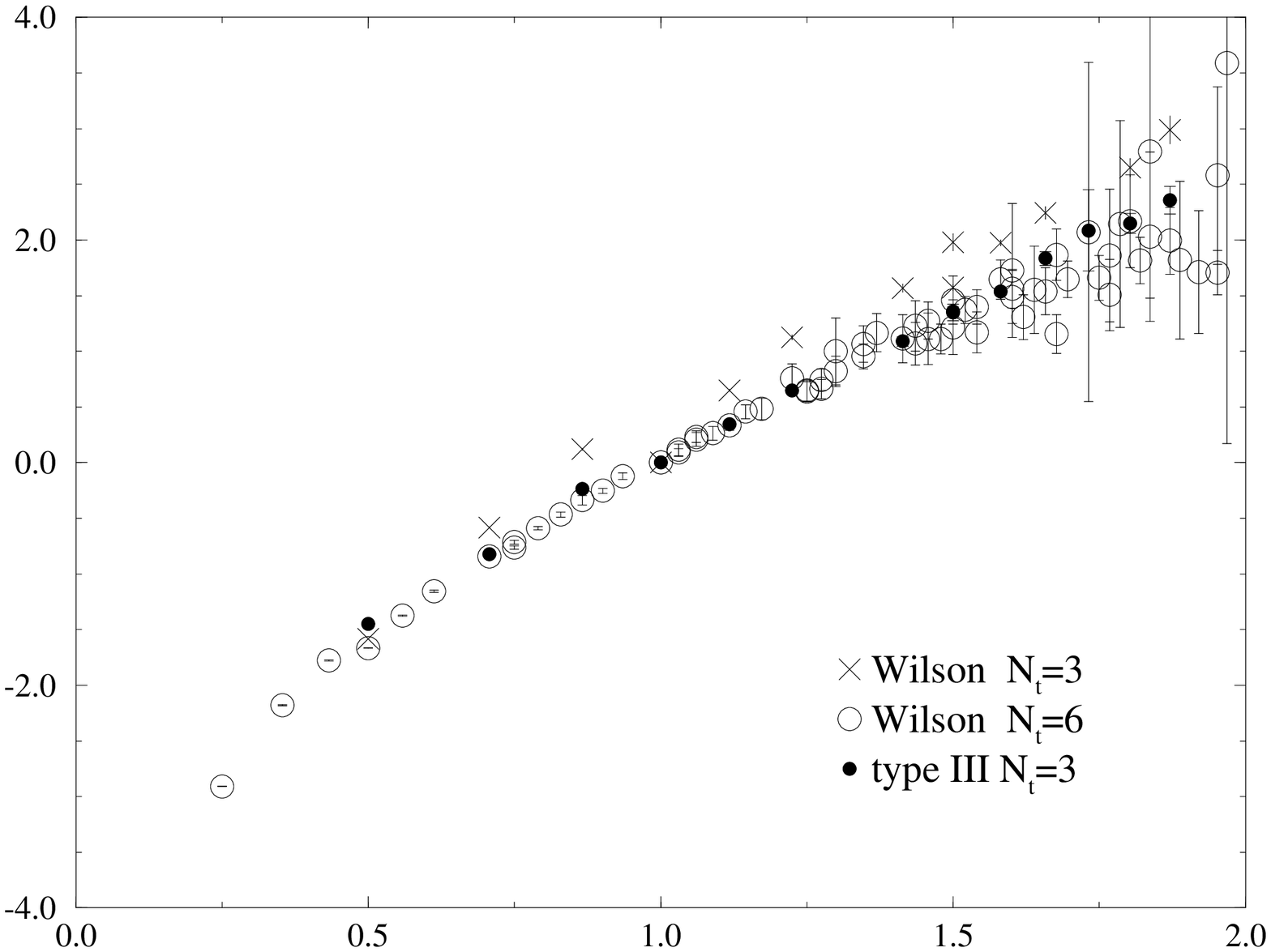}
\end{minipage}
\caption{ The $q\bar{q}$-potential $V(r;T)/T_c$ vs.\
$r T_c$ for different actions at $T=2/3\,T_c$.
Upper left: Wilson, upper right: type~I, lower left: type~III action.
These three plots correspond to $N_t=3$. 
The dashed line shows the function $-0.24/r + 2.48 r - 2.19$ ,
and serves to guide the eye.
The lower right graph shows again the potential for type~III and
Wilson action for $N_t=3$, together with the $N_t=6$ result for the 
Wilson action. The spatial size in all cases is $N_s=2N_t$.
}
\label{fig:perfect24}
\end{figure}

% =========================================================================
To repeat similar measurements of ref.~\cite{DEGRAND2} we have measured
the static $q\bar{q}$-potential $V(r;T)$ at finite temperature
$T=\frac{2}{3} T_c$ for the type~III action.  The result for
$\beta=\beta_c(N_t=2)=3.40$ on a lattice $6^3\times 3$ is shown on
fig.~\ref{fig:perfect24}.  The constant in the potential is fixed by
setting $V(r=1/T_c;T)=0$. For the Wilson action we used 
$12^3\times 6$ and $6^3\times 3$ lattices, again at $T=\frac{2}{3} T_c$.
These data and those for type~I action are taken from
ref.~\cite{DEGRAND2}. As opposed to the Wilson action, both FP actions
for $N_t=3$ have a proper slope at large distances. One also observes that
the type~III action has a somewhat smoother potential than type~I.   The
difference is, however, not as striking as for the perturbative
potential on fig.~\ref{fig:quadpot}. 
Note also a phenomenon which looks strange at first sight.  
The third point from the right on the type~I and type~III plots
 --- which corresponds to the diagonal distance (2,2,2) --- 
has larger error bars than the others. In the Wilson case
this point is not plotted because the measured correlation value was
lost in the noise.  This is also a sign of strong violation of rotational
symmetry, and is not completely restored in the
parametrized FP actions either. The reason for this phenomenon 
(besides the trivial factor of different multiplicities) is,
perhaps, that in the Wilson action there is no direct interaction
between diagonally separated links, and in the actual parametrization 
of the FP actions used here no spatial diagonal interaction term is 
included --- this would be an 8-link `parallelogram'.
(Such direct interaction term is certainly present in the true FP
action.)
As a result, the diagonally separated Polyakov loops are not bound
to each other so strongly as they should be.
A similar behaviour is observed for the distance (1,1,1), but the errors
are too small there to be seen in the figure.

% ======================================================================
\section{Conclusion}
\label{se:concl}
% ======================================================================
Using a more general smearing kernel we obtained a new FP action which
in the quadratic approximation performs better than those obtained
previously --- it has a somewhat faster decay rate for the interaction
coefficients and produces a perturbative potential close to the
continuum one. As far as the cut--off effects for finite $\beta$ are
concerned, the performance of this new FP action, or rather of the
actual parametrization suggested here, has to be investigated further.
To improve the parametrization it could also be useful to include 
rough classical solutions (like instantons or constant Abelian fields)
with analytically known action values \cite{BLATTER,BURKHALTER}.

As mentioned in the introduction, A.~Papa has measured the cut-off
effects in the free energy for type~I and type~IIIa actions \cite{PAPA}.
For the type~IIIa action at $T=2 T_c$ on a lattice with temporal
extension $N_t=3$ no cut-off effect has been observed within the small
errors --- the result agrees with the continuum prediction of
ref.~\cite{BEINLICH,BOYD2} and even at $N_t=2$ the error is 
$\sim 10\%$. 
For the type~I the deviation is $\sim 10\%$ at  $N_t=3$.
This has to be compared to the Wilson case where the cut--off effect
is $\sim 100\%$ at $N_t=2$ and even at $N_t=4$ it is still $\sim
20\%$.
Because the necessary simulation time grows very fast with $N_t$ --- 
about $N_t^{10}$ --- it is essential to keep the cut--off effects as
small as possible.

\medskip
\noindent
{\large \bf Acknowledgments:}

\noindent
We wish to thank Ruedi Burkhalter, Tom DeGrand, Anna Hasenfratz
and Alessandro Papa for useful discussions.
Especially, we would like to thank Peter Hasenfratz for many
helpful conversations and for reading the manuscript.  The codes not
specific to type~III RGT were taken over from previous work
\cite{DEGRAND1,DEGRAND2}.

% \newpage

% ========================================================================

\eject

\end{document}